\documentclass[12pt]{article}
\usepackage{slashed}
\usepackage{amsmath}
\usepackage{amssymb}
\usepackage{amsfonts}
\usepackage{color}
\usepackage{graphicx}
\newcommand{\nin}{\noindent}
\newcommand{\be}{\begin{equation}}
\newcommand{\ee}{\end{equation}}
\newcommand{\bea}{\begin{eqnarray}}
\newcommand{\eea}{\end{eqnarray}}

\newcommand{\nn}{\nonumber\\}

\newcommand{\ol}{\overline}

\begin{document}
 
\title{Consistent description of field theories with non-Hermitian mass terms}

\author{\underline{Jean Alexandre}${}^{a,}$\footnote{jean.alexandre@kcl.ac.uk}, Peter Millington${}^{b,}$\footnote{p.millington@nottingham.ac.uk}
and Dries Seynaeve${}^{a,}$\footnote{dries.seynaeve@kcl.ac.uk}\\
\small{${}^a$ Department of Physics, King's College London,}\\ \small{London WC2R 2LS, United Kingdom}\\ 
\small{${}^b$ School of Physics and Astronomy, University of Nottingham,}\\ \small{Nottingham NG7 2RD, United Kingdom}}
\date{}

\maketitle

\begin{abstract}
We review how to describe a field theory that includes a non-Hermitian mass term in the region of parameter space where the Lagrangian
is $\mathcal{PT}$-symmetric. The discrete symmetries of the system are essential for understanding the consistency of the model,
and the link between conserved current and variation of the Lagrangian has to be revisited in the case of continuous symmetries. 
\end{abstract}

\section{Introduction}

Among the potential extensions of the Standard Model is the possibility to consider non-Hermitian operators, as long as the corresponding energies 
remain real and the evolution remains unitary. It is known in Quantum Mechanics that a Hamiltonian that is symmetric under the combined action of parity ($\mathcal{P}$) and time 
reversal ($\mathcal{T}$) has real eigenvalues \cite{B}, and we 
review here some recent work done in scalar and fermionic field theories that involve non-Hermitian mass terms. In the region of parameter space where the
Lagrangian is $\mathcal{PT}$-symmetric, we show that a consistent description of such a system can be achieved if one associates the degrees of freedom with 
the $\mathcal{PT}$-conjugate fields, 
rather than the Hermitian-conjugate ones. After reviewing two different non-interacting models, we explain some features of a non-Hermitian extension of Quantum Electrodynamics.

\section{Anti-Hermitian mass terms}

The scalar and fermion models that we consider are described by the Lagrangians
\bea\label{Lsf}
\mathcal{L}_s &=&\begin{pmatrix}
\partial_\nu \phi_{1}^{\star} & \partial_\nu \phi_{2}^{\star}
\end{pmatrix}\begin{pmatrix}
\partial^\nu \phi_1 \\ \partial^\nu \phi_2
\end{pmatrix}
-\begin{pmatrix}
\phi_{1}^{\star} &  \phi_{2}^{\star}
\end{pmatrix} \begin{pmatrix}
m_1^2 & \mu^2 \\ 
- \mu^2 & m_2^2
\end{pmatrix} 
\begin{pmatrix}
\phi_1 \\  \phi_2
\end{pmatrix}~~~~\mbox{and}\nn
{\cal L}_f &=&\ol\psi\left(i\slashed\partial - m -\mu\gamma^5\right)\psi~,
\eea
and contain the anti-Hermitian terms $\mu^2(\phi_1^\star\phi_2-\phi_1\phi_2^\star)$ and $\mu\ol\psi\gamma^5\psi$, respectively.

In the scalar case \cite{AMS}, the eigenvalues of the mass matrix are
\be\label{eigenmassscalar}
M^2_{s\pm} =\frac{1}{2}(m_1^2+m_2^2) \pm \frac{1}{2}\sqrt{(m_1^2-m_2^2)^{2}-4\mu^{4}}~,
\ee
and these lead to real energies as long as $2\mu^2\leq|m_1^2-m_2^2|$.  
The corresponding eigenvectors are not orthogonal with respect to Hermitian conjugation and become parallel
in the limit where the eigenmasses become degenerate. At this degenerate point, the system describes half of the original degrees of freedom. 

In the fermionic case, originally introduced in \cite{BJR} and also related to \cite{JSM}, the energies are real as long as $|\mu|\leq m$, and the mass is 
\be
M_f=\sqrt{m^2-\mu^2}~.
\ee
It is interesting to see that, in the limits $\mu\to \pm m$ where the fermion is effectively massless, 
one loses either the right-chiral component or the left-chiral component. Indeed, in terms of  
\be
\psi_R=\frac{1}{2}(1+\gamma^5)\psi~~~~\mbox{and}~~~~\psi_L=\frac{1}{2}(1-\gamma^5)\psi~,
\ee
the Lagrangian reads
\be
{\cal L}_f=\ol\psi_R i\slashed\partial\psi_R+\ol\psi_L i\slashed\partial\psi_L-\ol\psi_L(m+\mu)\psi_R-\ol\psi_R(m-\mu)\psi_L~.
\ee
The limit $\mu=m$, for example, leads to the equation of motion $i\slashed\partial\psi_L=2m\psi_R$ and therefore ${\cal L}_f^{\rm on-shell}=\ol\psi_R i\slashed\partial\psi_R$.
Similarly, the limit $\mu=-m$ leads to ${\cal L}_f^{\rm on-shell}=\ol\psi_L i\slashed\partial\psi_L$.

In both the scalar and fermionic cases, one then reaches a singularity in some specific limits for $\mu$, since one loses half of the degrees of freedom. These ``exceptional points'' 
correspond to the boundaries, in parameter space, beyond which the $\mathcal{PT}$ symmetry is broken.

\section{$\mathcal{PT}$ symmetry and equations of motion}

A generic feature of non-Hermitian models with $\mathcal{PT}$ symmetry is their interpretation in terms of coupled systems with gain and loss. 
For the scalar model, one can see from the conserved current (see the next section) that one field plays the role of source and the other plays the role of sink; 
$\mathcal{P}$ swaps the fields, whereas $\mathcal{T}$ swaps gain and loss.
For this picture to be consistent, one field must transform as a pseudo-scalar \cite{AMS}, and the $\mathcal{PT}$-conjugate transpose of the scalar doublet is
\be
\Phi^\ddagger=\left[\begin{pmatrix} 1 & 0 \\ 0 & -1\end{pmatrix}\Phi^\star\right]^\mathsf{T}=(\phi_1^\star, -\phi_2^\star)
~~~~\mbox{where}~~~~\Phi=\begin{pmatrix}\phi_1 \\  \phi_2\end{pmatrix}~.
\ee
The scalar Lagrangian can then be written in an explicitly $\mathcal{PT}$-symmetric way as
\be\label{symmetricform}
{\cal L}_s = \Phi^{\ddagger}\begin{pmatrix} -\,\Box-m_1^2 & -\mu^2 \\ -\mu^2 & \Box+m_2^2 \end{pmatrix}\Phi~,
\ee
such that the equations of motion are obtained from the variational principle
\be\label{equamotscal}
\frac{\delta S_s}{\delta\Phi^{\ddagger}} = 0~~~~\text{or}~~~~ \left(\frac{\delta S_s}{\delta\Phi}\right)^{\ddagger}  = 0~,
\ee
where $S_s$ is the $\mathcal{PT}$-symmetric action.

The $\mathcal{PT}$-conjugate transpose of the fermion field is 
\be
\psi^\ddagger=[i\gamma^0\gamma^1\gamma^3\psi^\star]^\mathsf{T}=i\ol\psi\gamma^3\gamma^1~, 
\ee
and the explicitly $\mathcal{PT}$-symmetric form of the fermionic Lagrangian is
\be
{\cal L}_f =-i\psi^\ddagger(x)\gamma^1\gamma^3(i\overset{\leftrightarrow}{\slashed\partial}-m-\mu\gamma^5)\psi(x)~,
\ee
where 
\be
\overset{\leftrightarrow}{\slashed\partial}\equiv\frac{1}{2}\left(\overset{\rightarrow}{\slashed\partial}-\overset{\leftarrow}{\slashed\partial}\right)~.
\ee
The equations of motion are then obtained from the variational principle
\be
\frac{\delta S_f}{\delta\psi^\ddagger} = 0~~~~\mbox{or}~~~~ \left(\frac{\delta S_f}{\delta\psi}\right)^{\ddagger} = 0~.
\ee

Equivalent equations of motion for the scalar and the fermion models are obtained from the functional variations   
\bea
\mbox{(scalar)}~~~~~~~~&&\frac{\delta S_s}{\delta\Phi^\star} = 0~~~~\text{or}~~~~\frac{\delta S_s^\star}{\delta\Phi} = 0\\
\mbox{(fermion)}~~~~~~~&&\frac{\delta S_f}{\delta\ol\psi}=0~~~~\text{or}~~~~ \frac{\delta S_f^\star}{\delta\psi} = 0~.\nonumber
\eea
In addition, we note that one could instead choose the equations of motion to be defined via
\bea
\mbox{(scalar)}~~~~~~~~&&\frac{\delta S_s^\star}{\delta\Phi^\star} = 0~~~~\text{or}~~~~\frac{\delta S_s}{\delta\Phi} = 0\\
\mbox{(fermion)}~~~~~~~&&\frac{\delta S_f^\star}{\delta\ol\psi}=0~~~~\text{or}~~~~ \frac{\delta S_f}{\delta\psi} = 0~,\nonumber
\eea
which would correspond to the change $\mu^2\to-\mu^2$ (scalar case) or $\mu\to-\mu$ (fermion case). 
This would, however, not change the physical predictions, since it is equivalent to interchanging $\phi_1\leftrightarrow\phi_2$ for the scalar case and 
$\psi_R\leftrightarrow\psi_L$ for the fermionic case.

\section{Conserved currents}

By making use of the equations of motion, one can show that the conserved current for the scalar model is 
\be
j_s^\nu = i  \left(\phi_1^\star \partial^\nu \phi_1 - \phi_1 \partial^\nu \phi_1^\star\right)
-i\left(\phi_2^\star \partial^\nu \phi_2 - \phi_2 \partial^\nu\phi_2^\star\right)~,
\ee
which corresponds to the phase transformation \cite{AMS}
\be\label{phases}
\Phi' = \exp\left[+\,i\alpha \begin{pmatrix} 1 & 0 \\ 0 & -1 \end{pmatrix}\right]\Phi~~~~;~~~~
\Phi^{\ddagger\prime} = \Phi^{\ddagger}\exp\left[-\,i\alpha \begin{pmatrix} 1 & 0 \\ 0 & -1 \end{pmatrix}\right]~.
\ee
Notice that the sign difference between the components of the doublet reflects the source/sink behaviour. For the fermionic case, the conserved current is \cite{AB}
\be\label{conservedf}
j_f^\nu = \ol\psi\gamma^{\nu}\bigg(1+\frac{\mu}{m}\gamma^5\bigg)\psi~,
\ee
which corresponds to the phase transformation \cite{AMS}
\be\label{phasef}
\psi' = \exp\Big[+i\alpha\Big(1+\frac{\mu}{m}\gamma^5\Big)\Big]\psi~~~~;~~~~
\ol\psi'= \ol\psi\exp\Big[-i\alpha\Big(1-\frac{\mu}{m}\gamma^5\Big)\Big]~.
\ee
However, in both cases, the phase transformations (\ref{phases}) or (\ref{phasef}) do not leave the respective Lagrangian invariant, and the usual link between
invariance of the Lagrangian and current conservation does not hold \cite{AMS}.

The reason for this becomes clear when one considers the full structure of the variation of the Lagrangian under a field transformation, which in the scalar case reads
\be
\delta{\cal L}_s = \bigg(\frac{\partial \mathcal{L}_s}{\partial \Phi}-\partial_{\nu}\frac{\partial \mathcal{L}_s}{\partial(\partial_{\nu}\Phi)}\bigg)\delta\Phi
+\delta\Phi^{\ddagger}\bigg(\frac{\partial \mathcal{L}_s}{\partial \Phi^{\ddagger}}-\partial_{\nu}\frac{\partial \mathcal{L}_s}{\partial(\partial_{\nu}\Phi^{\ddagger})}\bigg)
+\partial_\nu (\delta j^\nu)~.
\ee
Since the theory is not Hermitian, the two parentheses on the right-hand side of the previous equation cannot simultaneously vanish on-shell. If we 
choose the equations of motion to be defined by eq.(\ref{equamotscal}) then current conservation requires
\be\label{conds}
\delta{\cal L}_s = \bigg(\frac{\partial \mathcal{L}_s}{\partial \Phi}-\partial_{\nu}\frac{\partial \mathcal{L}_s}{\partial(\partial_{\nu}\Phi)}\bigg)\delta\Phi
=2\mu^2(\phi_2^\star\delta\phi_1-\phi_1^\star\delta\phi_2)~.
\ee
Similarly, current conservation in the fermionic case requires that the Lagrangian variation is
\be\label{condf}
\delta{\cal L}_f = \bigg(\frac{\partial{\cal L}_f}{\partial\psi}-\partial_\nu\frac{\partial{\cal L}_f}{\partial(\partial_\nu\psi)}\bigg)\delta\psi =-2\mu\ol\psi\gamma^5\delta\psi~.
\ee
One cannot therefore have a conserved current together with invariance of the non-Hermitian part of the Lagrangian.  Current conservation nevertheless remains 
the essential physical feature, such that invariance of the Lagrangian is not required, as it would be for an Hermitian theory. 
Instead, the required transformation of the Lagrangian is fixed by an identity of the form (\ref{conds}) or (\ref{condf}), 
and, in the next section, we give 
an alternative derivation for the fermionic conserved current, which does not rely on the variational procedure \cite{AMS}.

\section{Non-unitary map}

The fermionic equation of motion can be written in the Schr\"odinger form 
\be
i\partial_0\psi=\gamma_0(\vec\gamma\cdot\vec p+m+\mu\gamma^5)\psi~,
\ee
and one can look for a map $\chi=B\psi$ that leads to 
\be
i\partial_0\chi=\gamma_0(\vec\gamma\cdot\vec p+M_f)\chi~~~~\mbox{with}~~M_f=\sqrt{m^2-\mu^2}~.
\ee
The latter equation of motion arises from the Hermitian Lagrangian
\be
{\cal L}_\chi=\ol\chi(i\slashed\partial-M_f)\chi~,
\ee
for which we know that the $U(1)$ conserved current is $j_f^\nu=\ol\chi\gamma^\nu\chi$. Expressed in terms of the original field, the current is then 
\be\label{conservedfbis}
j_f^\nu=\ol\psi\gamma^0B^\dagger\gamma^0\gamma^\nu B\psi~.
\ee

The transformation matrix $B$ is not unitary, since it maps an equation of motion obtained from a non-Hermitian Lagrangian 
to an equation of motion obtained from a Hermitian Lagrangian. We can fix the form of $B$, since it must satisfy
\be
B\gamma^0(\vec\gamma\cdot\vec p+m+\mu\gamma^5)B^{-1}=\gamma^0(\vec\gamma\cdot\vec p+M_f)~,
\ee
for any momentum $\vec p$, and it is easy to find that
\be
B\propto1+\gamma^5\sqrt\frac{1-\sqrt{1-\mu^2/m^2}}{1+\sqrt{1-\mu^2/m^2}}~.
\ee
Together with eq.~(\ref{conservedfbis}), we recover the conserved current (\ref{conservedf}). Note that the singularity of the limit $\mu^2\to m^2$ can be seen here
because $B$ becomes proportional to the projector $1+\gamma^5$ and thus has no inverse.

\section{Non-Hermitian extension of QED}

The gauged model for the fermionic Lagrangian is \cite{ABM}
\be\label{gaugedmodel}
{\cal L} =-\frac{1}{4}F^{\rho\sigma}F_{\rho\sigma}+\ol\psi\left[i\slashed\partial-\slashed A(g_V+g_A\gamma^5)-m-\mu\gamma^5 \right]\psi~,
\ee
where $F_{\rho\sigma}=\partial_\rho A_\sigma-\partial_\sigma A_\rho$. 
In the massless case ($m=\mu=0$), 
the action is invariant under the combined vector plus axial-vector gauge transformations
\bea
A_\rho &\longrightarrow& A_\rho' = A_\rho-\partial_\rho\theta\\
\psi &\longrightarrow& \psi' = \exp\left[i\left(g_V+g_A\gamma^5\right)\theta\right]\psi\nn
\ol\psi &\longrightarrow& \ol\psi^\prime = \ol\psi\exp\left[i\left(-g_V+g_A\gamma^5\right)\theta\right]~.\nonumber
\eea
For $0<|\mu|\le m$, the one-loop corrections in dimension $4-2\epsilon$ can be calculated in the chiral basis, and the divergent part of the polarisation tensor is \cite{ABM}
\be
\Pi^{\rho\sigma}=\frac{g_V^2+g_A^2}{12\pi^2\epsilon}(p^\rho p^\sigma-p^2\eta^{\rho\sigma})+\frac{g_A^2}{\pi^2\epsilon}(m^2-\mu^2)\eta^{\rho\sigma}~,
\ee
becoming transverse when $\mu^2\to m^2$. In this limit, the full vector plus axial-vector
gauge symmetry is recovered, and this is consistent with our earlier observation (in section 2) that the theory effectively
becomes massless.

This model may have interesting implications for neutrino physics \cite{ABM}, since the probability density that one obtains from the current (\ref{conservedf}) is
\be
j^0=\left(1+\frac{\mu}{m}\right)|\psi_R|^2+\left(1-\frac{\mu}{m}\right)|\psi_L|^2~.
\ee
Thus, the contribution from the right-handed component, as well as the fermion mass, can be made arbitrarily small by choosing the ratio $\mu/m\simeq-1$ . 
Note that this result is consistent with the  
study \cite{Chernodub} of a non-Hermitian system of fermions on the lattice, where the numbers of right- and left-handed fermions are not the same. 
A more detailed description of a non-Hermitian gauge-Yukawa model is given in \cite{ABM} (see also \cite{ABM2}).

\section{Conclusion}

As shown here, the description of a non-Hermitian field theory that is $\mathcal{PT}$-symmetric is consistent if one considers the $\mathcal{PT}$-conjugate fields to be 
the relevant degrees of freedom, instead of the Hermitian-conjugate 
fields. Further studies of interacting theories, involving non-perturbative tools, are now being considered, specifically the Schwinger-Dyson or the Wilsonian approaches. 
For this, the consistency of the path-integral
quantisation necessarily requires us to integrate over $\mathcal{PT}$-conjugate pairs of degrees of freedom, consistent with the variational procedures described above.\\

\nin{\bf Aknowledgements} The work of PM is supported by STFC Grant No. ST/L000393/1 and a Leverhulme Trust Research Leadership Award.

\end{document}